\documentclass[a4paper,aps,prl,twocolumn,floatfix,showpacs,superscriptaddress,footnoteinbib]{revtex4-1}
\usepackage{graphics}
\usepackage{epsfig}
\usepackage{times}
\usepackage{xcolor}   
\usepackage{amsfonts}
\usepackage{amsmath}
\usepackage{amssymb}
\usepackage{amsthm}
\usepackage{hyperref} 
\usepackage{wasysym}
\usepackage{bm}

\newcommand{\sgn}{\mathop{\mathrm{sgn}}}

\newcommand{\overbar}[1]{\mkern 1.5mu\overline{\mkern-1.5mu#1\mkern-1.5mu}\mkern 1.5mu}

\begin{document}

\title{
Quantum Hall realization of polarized intensity interferometry}

\author{Krishanu Roychowdhury}
\affiliation{LASSP,  Department  of  Physics,  Cornell  University,  Ithaca,  NY  14853, USA}
\author{Disha Wadhawan}
\affiliation{Department of Physics and Astrophysics, University of Delhi, Delhi 110007, India}
\author{Poonam Mehta}
\affiliation{School of Physical Sciences, Jawaharlal Nehru University, New Delhi 110067, India}
\author{Biswajit Karmakar}
\affiliation{Saha Institute of Nuclear Physics, Bidhan nagar, Kolkata, West Bengal 700064, India} 
\author{Sourin Das}
\affiliation{Department of Physics and Astrophysics, University of Delhi, Delhi 110007, India}

\begin{abstract}
We combine the ideas of intensity interferometry, polarization optics and Bell's measurement into an experimental proposal which is hosted in a $\nu\,$=$\,2$ quantum Hall (QH) edge state. Our interferometer comprises of a single gate, that separates the spin resolved edge states of $\nu$\,=\,$2$ state. An analog of waveplate (from polarization optics) is realized by exposing the individual edges  to nano-magnets over a finite length which facilitates coherent manipulation of the electron spin on the edge. We show that the in-plane rotation of magnetization direction of the nano-magnets results in oscillations observed solely in the cross-correlated noise  which arises from two particle interference. Through numerical  simulations we demonstrate that our proposal is well within the reach of recent experimental developments of spin manipulations on the QH edge.      
\end{abstract}
\maketitle
\begin{figure*}
\begin{centering}
\includegraphics[width=1.0\linewidth]{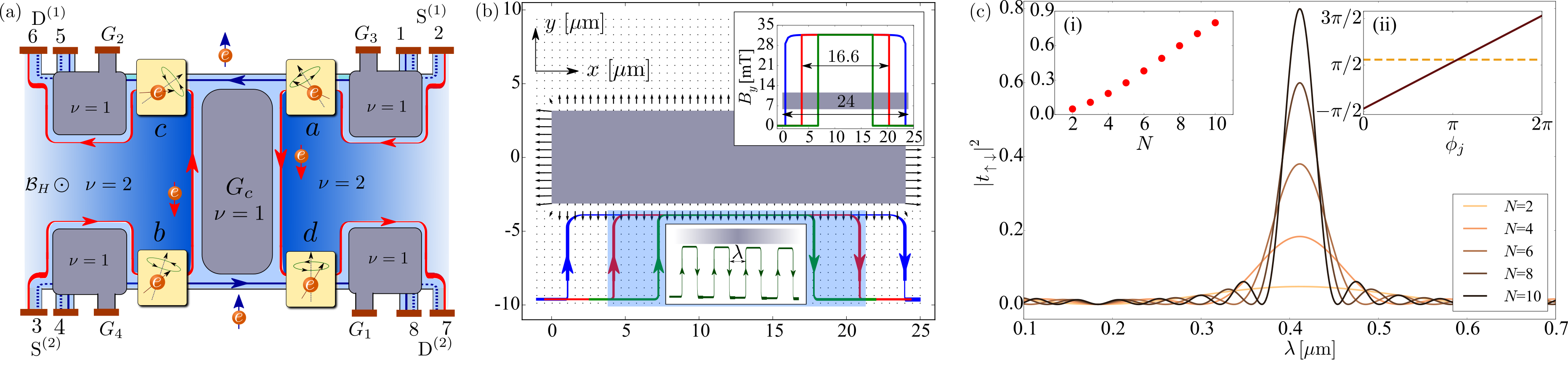}              
\end{centering}
\caption{(Color online) (a) Schematic of our proposed two-particle interferometer. Four magnetic mixers $a$,$b$,$c$,$d$ provide the tuning of phase and amplitude of spin mixing matrix elements that results in the oscillations in the cross-correlated noise measured between the detectors $D^{(1)}$ (5$\uparrow$ or 6$\downarrow$) and $D^{(2)}$ (7$\downarrow$ or 8$\uparrow$). Spin injection is facilitated by the sources $S^{(1)}$ (1$\uparrow$ or 2$\downarrow$ with a bias $V$) and $S^{(2)}$ (3$\downarrow$ or 4$\uparrow$ with a bias $V$. The edge channel for each of the spin species can be contacted separately using the gates $G_1$, $G_2$, $G_3$, and $G_4$. The middle gate $G_c$ is employed to maintain spatial separation between the edges and reduce the interactions away from the mixers. (b) Vector plot of the in-plane magnetic field produced by the bar magnet on the plane of the 2DEG.  The red, green, and blue curves represent possible paths of the $\nu=2$ edge to be taken in order to experience the field profile modeled by the function $\alpha(x)$. The upper inset shows variation of $\vert B_y\vert$ along these paths where the colors are in correspondence. 
(c) This figure shows resonant enhancement of spin transfer ensured by the resonance condition ($\lambda=\lambda_{\rm res}=\pi v_{\rm F}/\Delta$ with $\lambda_{\rm res}\approx0.41\mu$m) \cite{karmakar2011controlled}. Inset (i): at the resonance, the net phase of the off-diagonal element of $S_j^N$ is plotted as a function of  $\phi_j$. It shows that apart from an offset (constant) they are the same (plot for $N=6$ case). Inset(ii): it shows that absolute square of the net spin mixing amplitude ($t_{\uparrow\downarrow}$) grows monotonically with $N$ irrespective of the values of $\phi_j$'s in feasible range for $N$. 
\label{fig:magprof}}
\end{figure*}

%
\noindent
{\underline{\sl {Introduction:-}}}
\label{intro}
In nineteen fifties, Hanbury Brown and Twiss (HB-T) observed a subtle effect in intensity correlations of light coming from distant stars which was  used to measure the angular diameter of the stars~\cite{brown1956test}. This observation led to the birth of intensity interferometry. Since then intensity interferometry has been explored in a variety of fields ranging from nuclear to condensed matter physics and new perspectives are still being explored\cite{cotler15a,cotler15b}. Quantum mechanically, intensity interferometry refers to a phenomenon wherein a quantum state describing a pair of particles (emanating from two independent sources) interfere with itself.
Significant technological advancements in the past two decades have made it possible to study some of the interesting aspects of intensity interferometry in the context of electronic solid state devices. One of the first theoretical proposals in this context (based on QH edge states) was by Samuelsson et al.~\cite{samuelsson2004two} where orbital part of the electron wavefunction played a crucial role while spin part was frozen.
In context of optics, recently Mehta et. al~\cite{mehta10} proposed that exploiting the full vector nature of light (photon spin) in a HB-T set-up could give rise to non-trivial realization of multiparticle interference which was experimentally demonstrated for both classical light~\cite{satapathy12} and for photons~\cite{martin}.  
Coming back to the QH realization of the HB-T interferometer, manipulation of the spin in a sense similar to manipulation of polarization state of light by a  waveplate has not been explored so far. The primary reason has been the challenge in realizing a QH  counterpart of the waveplate. In this proposal, we theoretically explore a scenario where controlled rotation of electronic spin (analog of waveplate) is exploited, thereby, expanding the scope of HB-T interferometry and opening up potentially new avenues for manipulating the HB-T correlations, going beyond the proposals in the past~\cite{samuelsson2004two,neder2007interference,giovannetti2008multichannel}\footnote{It should be noted that Ref~[{\onlinecite{giovannetti2008multichannel}}] does discuss a possibility of mixing of copropagating edges which can naturally involve spin mixing but they do not explore the potential of such mixing for  generating interference which is the focus of our work. For Ref~[{\onlinecite{giovannetti2008multichannel}}] the mixing just facilitates formation of an AB loop and the AB flux is used to generate the interference which is not the case for us}. All the earlier proposals used orbital degree of freedom via AB flux,  while we exploit coherent spin rotation in order to produce the HB-T interference pattern.  We also study control of the orbital entanglement generated in our proposed set up via spin manipulations and its interesting consequence on visibility. The experimental feasibility of our proposal relies solely on the recent developments reported by Karmarkar et al.~\cite{karmakar2011controlled} where nano-magnets were deposited locally on the edge  to rotate the spin of an electron living on a composite $\nu=2$ edge state~\cite{muller92}. Moreover, spin-resolved injection and detection of the electrons are also crucial for the proposal which is routinely done in experiments on $\nu=2$ states~\cite{karmakar2011controlled, karmakar2015controlled}.  \\
\noindent
{\underline{\sl{Set-up and Model Hamiltonian:-}}}
\label{model}
A schematic of the proposed set-up is shown in Fig.~\ref{fig:magprof}(a). The figure represents a minimal scenario for a two source ($S^{(1,2)}$) -- two detector ($D^{(1,2)}$) set-up where an electron shot from a given source has a finite probability of reaching either of the two detectors which is an essential requirement for realizing HB-T interferometry. Our set-up comprises of a Hall bar geometry hosting a $\nu=2$ state with a single gate in the middle which is tuned such that the inner edge of the $\nu=2$ state is fully reflecting while the outer edge is fully transmitting. Note that the inner and outer edge states are spin polarized with opposite polarization (down and up) with respect to the direction of the quantizing magnetic field ($\mathcal{B}_H$) responsible for the QH state. Additionally local in-plane magnetic fields are applied in four patches (call them mixer) on the edge-- $a$,$b$,$c$,$d$ in Fig.~\ref{fig:magprof}(a) which leads to coherent mixing of the spin up and spin down edge states as they impinge on the mixers hence making them an analog of a waveplate in optics. 
Also note that the local mixing of edge states has been achieved experimentally in Ref.~\cite{karmakar2011controlled} and the coherence of such mixing has also been established recently in Ref.~\cite{karmakar2015controlled}. It should be noted that the copropagating spin up and spin down edge states are in close physical vicinity of each other and hence effects due to local inter electron interactions are expected to influence our results~\cite{degiovanni2010plasmon, bocquillon2013separation, chirolli2013interactions, wahl2014interactions}. To avoid such complication we have proposed a geometry where the copropagating  edge states are separated out in space from each other using gates to strongly suppress the effects due to interaction  except in the mixer region [see Fig.~\ref{fig:magprof}(a)]. Later we have shown that the interactions in the mixer regions do not modify our results in experimentally feasible window of parameter space. The four continuous edges (two spin up and two spin down) which connect $S^{(1,2)}$ to $D^{(1,2)}$ are described by the model Hamiltonian~\cite{halperin82} given by 
\begin{equation}
 \mathcal{H}_{0} = \int_{-\infty}^{\infty} dx ~ \psi_{\sigma}^{\dagger}\{-i v_{\text{\textit{F}}}\partial_{x}+\sgn({\sigma})\Delta\}\psi_{\sigma} ,
 \label{h_free}
\end{equation}
and that for each of the mixers is described by
\begin{equation}
 \mathcal{H}_{B} = \int_{-\infty}^{\infty} dx ~  \alpha(x) \left\{(\mathcal{B}/2)\psi_{\sigma}^{\dagger} \psi_{\overbar{\sigma}} + \text{h.c.} \right\}.
 \label{h_mix}
\end{equation}
Here  $x$ is an intrinsic coordinate associated with each continuous edge. In writing $\mathcal{H}_{0}$ and $\mathcal{H}_{B}$, we have assumed $\hbar=\mu_B=g=1$ ($g$ being the Land\'{e} $g$-factor). The $z$-component of the electron spin is given by $\sigma$ and its complementary by $\overbar{\sigma}$. The in-plane magnetic mixer is placed over a length of $l$ expressed by the function $\alpha(x)=\Theta(x+l/2)-\Theta(x-l/2)$ and the parameter $\Delta$ denotes the momentum mismatch ($k_{\uparrow}-k_{\downarrow} = 2\Delta/v_{\text{F}}$) between the spin channels~\cite{karmakar2011controlled}. The quantity $\mathcal{B}=2\Gamma e^{i\phi}$ represents effect of the local in-plane magnetic field of strength $2\Gamma$ with a relative orientation of an angle $\phi$ with respect to the $X$-axis of a coordinate system which is chosen such that the 2DEG hosting the QH state lies on the $X$-$Y$ plane. 

The scattering matrix $S_{j}$ for a given mixer $j$ [$j=a,b,c,d$ as in Fig.~\ref{fig:magprof}(a)] which connects the incoming spin channels to the outgoing ones, takes the form
\begin{equation}
S_{j} =\begin{pmatrix}
  \langle\uparrow\vert\hat{S_{j}}\vert\uparrow\rangle & \langle\uparrow\vert\hat{S_{j}}\vert\downarrow\rangle \\
  \langle\downarrow\vert\hat{S_{j}}\vert\uparrow\rangle & \langle\downarrow\vert\hat{S_{j}}\vert\downarrow\rangle
 \end{pmatrix}~=\begin{pmatrix}
  A_j & -iB_j^{*} \\
  -iB_j & A_j^{*} 
 \end{pmatrix},
\label{s_mat1}
\end{equation}
where 
\begin{align}\label{params}
 A_j &= e^{-i\Delta l_j/v_{\text{F}}}\Bigg(\cos{\frac{E_{0}l_j}{v_{\text{F}}}}+i \frac{\Delta}{E_{0}} \sin{\frac{E_{0} l_j}{v_{\text{F}}}}\Bigg) \equiv \vert A_j\vert e^{-i\tilde{\phi}_A}, \nonumber\\
 B_j &= e^{-i\Delta l_j/v_{\text{F}}}\frac{\Gamma_j e^{-i\phi_j}}{E_{0}}\sin{\frac{E_{0}l_j}{v_{\text{F}}}} \equiv \vert B_j\vert e^{-i\phi_j}e^{-i\Delta l_j/v_{\text{F}}}, \nonumber\\
 E_{0} &= \sqrt{\Delta^2+\Gamma_j^2}.
\end{align}
Here $j$ is the mixer index which runs over $j=a,b,c,d$.  Note that the diagonal elements denote the spin conserving amplitude while the off-diagonals represent spin mixing amplitudes. Using unitarity of the scattering matrix, we further parametrize $\vert A_j\vert\equiv\cos(\theta/2)$ and $\vert B_j\vert\equiv\sin(\theta/2)$ which provides a direct geometric interpretation of the action of the mixers on spin states. When an electron is incident on the mixer from the spin up (down) edge, the outgoing electron obtains a spin orientation whose polar angle is $\theta$ ($\pi-\theta$) and azimuthal angle is  $\pi/2+\phi_j-\tilde{\phi}_A$ ($3\pi/2+\phi_j-\tilde{\phi}_A$). In general, all $l_j$ and $\Gamma_j$ can be different, but we assume them to be equal for algebraic simplification and henceforth denoted them by  $l$ and $\Gamma$ respectively. This implies that all $A_j$ are equal and hence forth  we denote them by $A$.  So the scattering matrix for all the four magnetic mixers differ only in the phase $\phi_j$ which can be tuned by choosing different in-plane orientations of the magnetic field in each of the four mixers which is the  geometric degree of freedom not been explored before. \\
Now we will proceed to show that this in-plane rotation of the magnetic field, {\it viz.} tuning $\phi$ with $\Gamma$ kept constant, in any one of the mixers leads to oscillations only in the cross-correlated noise measured between two drains ($D^{(1)}$ and $D^{(2)}$) while the current and the auto-correlated noise measured at each drain remain insensitive to it. These oscillations are manifestations of two-particle interference pattern~\cite{blanter2000shot,neder2007entanglement, moskalets2016single}. Producing these oscillations with spin manipulations alone constitutes the central result of this rapid communication. Our proposal, thus, provides the first analog of {\it HB-T interferometry using different polarized states of light} in a QH set-up with a neat geometric control parameter for tuning the two-particle interference pattern. \\
\noindent
{\underline{\sl{Cross-correlated noise:-}}}
\label{noise}
The proposed set-up in Fig.~\ref{fig:magprof}(a) shows two sources $S^{(1)}$ and $S^{(2)}$ that are held at a bias $V$ with respect to the detectors $D^{(1)}$ and $D^{(2)}$ (both being  virtually grounded) between which the cross-correlated noise is measured~\footnote{It should be noted that the two oppositely spin polarized edge constituting the composite edge of $\nu=2$ can be contacted separately in realistic experiments \cite{karmakar2011controlled, karmakar2015controlled}. The individual constituent of the composite $\nu=2$ edges being spin polarized, these contacts naturally provide a spin-resolved injection at the source and spin-resolved measurement in the detectors.}. Following Ref.~\cite{blanter2000shot} we can express the spin resolved cross-correlated noise between detectors $D^{(1)}$ and $D^{(2)}$ at zero temperature as
\begin{equation}
 S^{\gamma\delta}_{\alpha\beta} = f' \left( \vert s_{\alpha\gamma}\vert^{2}\vert s_{\beta\gamma}\vert^{2}+\vert s_{\alpha\delta}\vert^{2} \vert s_{\beta\delta}\vert^{2}+2 \mathcal{R}e [s_{\beta\gamma}s_{\alpha\gamma}^{*}s_{\alpha\delta}s_{\beta\delta}^{*}] \right).              
\label{noise_1}
\end{equation}
Here $\alpha,\beta$ correspond to the spin polarizations (up or down) of the electrons in the composite edge entering the drains while $\gamma$,$\delta$ correspond to spin polarizations of the electron in the composite edge emanating out of the two sources and $f'=-e^{3}V/\pi$~\footnote{One should note that once the polarization of the spin is selectively fixed both in the sources and the detectors, it makes a unique choice for all the indices, thus, no sum is implied in Eq.~(\ref{noise_1}).}. A typical term $s_{\alpha\gamma}$ represents the propagation amplitude of an electron starting from the source with spin polarization $\gamma$ going to the drain with spin polarization $\alpha$. It contains the elements of the respective $S$-matrices corresponding to the mixers that appear on their way. The last term in Eq.~(\ref{noise_1}) is the two-particle interference term~\cite{blanter2000shot}, henceforth denoted as $Z$. Using Eq.~(\ref{s_mat1}) and Eq.~(\ref{params}), this can be re-expressed as $Z = \tilde{\sigma}(\sin^4\theta \cos\zeta)/8$ where $\zeta\equiv (\phi_a+\phi_b)-(\phi_c+\phi_d)-4\tilde{\phi}_A + \phi_D+ \phi_{AB}$, $\phi_D$ being the sum of dynamical phase accumulated from free propagation and $\phi_{AB}$ is the net Aharonov-Bohm phase accumulated from the $\nu=1$ region below the single gate [see Fig. 1 (a)]. The factor $\tilde{\sigma}\equiv\sigma_{\alpha}\sigma_{\beta}\sigma_{\gamma}\sigma_{\delta}$ is the product of spin polarizations for the sources ($\gamma$,$\delta$) and the detectors ($\alpha$,$\beta$), each of them being $\pm1$ for up and down polarizations respectively. Now we discuss possibility of two distinct cases: (i) when both the detectors are detecting same spin species (i.e. $\tilde{\sigma}=\sigma_{\gamma}\sigma_{\delta}\equiv\sigma_0$) and (ii) when the detectors are detecting different spin species (i.e. $\tilde{\sigma}=-\sigma_0$). In case (i), the full noise expression $S^{\gamma\delta}_{\alpha\beta}$  reduces to 
\begin{equation}
 S_1 = \frac{f'}{8}[1+\sigma_0\cos\zeta]\sin^4\theta
 \label{noise_same}
\end{equation}
while in case (ii), it is $S_2 = \frac{f'}{4} \sin^2\theta - S_1$. Hence both $S_1$ and $S_2$ exhibit oscillations via the $\cos\zeta$ term as we vary any of the $\phi_j$'s from zero to $2\pi$. It is straight forward to check that both the average current and the auto-correlated noise evaluated for drains $D^{(1)}, D^{(2)}$ are independent of any $\phi_j$, hence, do not show any such oscillation. This establishes $\phi_j$'s as neat control parameters for producing the two-particle interference pattern. Note that when the spin mixing probability is 50 \% ($\theta=\pi/2$) the expressions for $S_1$ and $S_2$ can be written as a single one, given by  $S^{\gamma\delta}_{\alpha\beta}=\frac{f'}{8}[1+\tilde{\sigma}\cos\zeta]$~\footnote{a similar form appears in Ref.~\cite{chung2005visibility}}. Owing to the spin degrees of freedom the visibility, $\mathcal{V}=(S^{\gamma\delta}_{\alpha\beta}\vert_{\rm{max}}-S^{\gamma\delta}_{\alpha\beta}\vert_{\rm{min}})/ (S^{\gamma\delta}_{\alpha\beta}\vert_{\rm{max}}+S^{\gamma\delta}_{\alpha\beta}\vert_{\rm{min}})$ also bears a crucial difference between case (i) and (ii). While for case (i), it is fixed to 1~\cite{chung2005visibility}, a tunability in terms of $\theta$ remains in case (ii) given by $\mathcal{V}=(1-\cos^2\theta)/(1+\cos^2\theta)$. Here $S^{\gamma\delta}_{\alpha\beta}$ is extremized with respect to the interference phase $\zeta$ for obtaining $\mathcal{V}$. 

\noindent
{\underline{\sl{Entanglement:-}}} Though our interferometer does not produce entangled pairs of electron but a measurement of the cross correlated noise actually lead to a measurement of orbital entanglement which is a very interest aspect of such a set up. A typical two electron state entering the interferometer from sources (contact $1 \in S^{(1)}$ and contact $4 \in S^{(2)}$) can be identified as $|\Psi_{\text{in}}\rangle = \prod_{E \leq eV} c_1^{\dagger}(E) c_4^{\dagger}(E) \vert0\rangle$.  The corresponding outgoing state before the respective electrons enter mixer $c$ and $d$ is given by $|\Psi_{\text{out}}\rangle = \Big(A^* c_{c1}^{\dagger} + i B_a^* c_{d1}^{\dagger}\Big)\Big( A^* c_{c4}^{\dagger} + i B_b^* c_{d4}^{\dagger}\Big)\vert0\rangle$, where $c_{c1}^{\dagger}$ denotes the creation operator for an electron emitted from contact 1 and transmitted through mixer $a$ but yet to impinge on mixer $c$. Note that a part of the total out going product wave function $|\Psi_{\text{out}}\rangle$ consists of $c_{c1}^{\dagger}c_{d4}^{\dagger}$  and $c_{d1}^{\dagger}c_{c4}^{\dagger}$, representing a entangled wavefunction in the orbital labels 1 and 4  given by
\begin{equation}
|\Psi_{\text{cd}}\rangle = i\frac{A^*}{N}\Big( B_b^*c_{c1}^{\dagger}c_{d4}^{\dagger} - B_a^*c_{c4}^{\dagger}c_{d1}^{\dagger} \Big)\vert0\rangle,
 \label{out_2}
\end{equation}
where $N$ is the normalization constant. The cross-correlated noise measured between contacts 1 and 4 is actually a measure of this entanglement. Following Ref.~\cite{samuelsson2003orbital} the entanglement  can be quantified as the CHSH inequality~\cite{clauser1969proposed}  expressed in terms of zero frequency cross-correlated noise. Note that the cross-correlated noise represents the joint probabilities for two electrons to be detected in the two drain contacts simultaneously~\cite{neder2007interference}\footnote{It should be noted that the joint probabilities remain finite even in absence of one of the sources, which reduces it to a Y-junction geometry, but the interference term does not appear in the cross correlated noise}. In our set-up four such joint probabilities can be defined between the drain contacts $D^{(1)}$ and $D^{(2)}$ i.e, contact $5$-$8$ as $P_{\uparrow\uparrow}(=S_1/S_0$), $6$-$7$ as $P_{\downarrow\downarrow}(=S_1/S_0$), $5$-$7$ as $P_{\uparrow\downarrow}(=S_2/S_0$) and $6$-$8$ as $P_{\downarrow\uparrow}(=S_2/S_0$) where $S_0=(f'/2)\sin^2\theta$ is the sum of all spin resolved cross-correlated noises measured between $D^{(1)}$ and $D^{(2)}$. The CHSH inequality is formulated in terms of $E(\phi_c,\phi_d) = P_{\uparrow\uparrow}+P_{\downarrow\downarrow}-P_{\uparrow\downarrow}-P_{\downarrow\uparrow}$. In particular for 50\% probability of spin mixing at all mixers, $E(\phi_c,\phi_d)=\cos\zeta$ and the CHSH inequality  reduces to $0\leq\vert{\cos\zeta(\phi_c,\phi_d)-\cos\zeta(\phi_c,\phi'_d)+\cos\zeta(\phi'_c,\phi_d)+\cos\zeta(\phi'_c,\phi'_d)}\\
\vert \leq2$. Note that the maximally entangled states lead to  violation of this inequality.\\
The initial condition required to prepare a maximally entangled state in our set-up corresponds to $\phi_a=\phi_b=0$ [from Eq.~(\ref{out_2})]. Further measurements involving mixers $c$ and $d$ can now lead to a maximal violation of the inequality if we tune to $\phi_c=\phi_0-\pi/4$; $\phi_d=\pi/2$; $\phi'_c=\phi_0-3\pi/4$; $\phi'_d=\pi$ (where $\phi_0\equiv\phi_D+\phi_{AB}-4\tilde{\phi}_A$) while $\theta=\pi/2$. Hence, in principle our set-up allows for a measurement where maximal violation of CHSH inequality could be observed.\\ 
Note that the feasibility of observing the violation of CHSH inequality and/or the oscillations in noise hinges upon three main points:  (1) can we realize a step function like magnetic field profile given in Eq.~(\ref{h_mix}) (which leads to the neat $\phi_{j}$ dependence in scattering matrix $S_{j}$) in a realistic set-up using nano-magnets (2) how much control do we have on parameter $\theta$ and (3) how can one rotate the direction of the in-plane magnetic fields (i.e. control $\phi_j$) acting on the composite edge in a desired fashion. Regarding point (3), a control of $\phi_{j}$ can be generated if one deposits multiple nano-magnets with different orientation and uses gates to redirect the path of edge states to access these different orientations. Rest of the points are discussed in detail below. 
%

%
%
\noindent
{\underline{\sl{Feasibility study:-}}}
\label{expt}
To address point (1) raised above, we perform a simulation following Ref.~\cite{karmakar2011controlled} for finding the in-plane components of the magnetic field produced by a nano-magnet which is placed above the 2DEG. The simulation is aided by a model of bar magnet which is $24\mu$m long, $7\mu$m wide, $120$nm thick, and is placed $100$nm above the 2DEG. Results of the numerics are summarized in Fig.~\ref{fig:magprof} (b) and (c). The primary task now is to confirm that there are paths in the 2DEG on which the magnetic field strength mimics a step function like profile as given in Eq.~(\ref{h_mix}). Once this turns out to be the case, one can use electrical gates to deform the edge states to guide their motion on those paths. Fig.~\ref{fig:magprof}(b) shows a vector plot of the in-plane components of the field and the upper inset shows variation of $y$-component of the magnetic field ($B_y$) along the three physical paths in red, blue and green in the main panel. It is clear from the figure that the paths in green and red actually do produce the desired step function like profile for the magnetic field on a nanometer scale resolution. We have plotted the $y$-component alone in the inset as the $x$-component is negligible with respect to the $y$-component on the green and the red paths and hence of no consequence. The $z$-component has the effect of renormalizing $\Delta$ and it does not influence the $\phi_j$ dependence in corresponding scattering matrix $S_j$, hence, can be neglected . Note that if electrostatic gates are to be used to guide the edge states on a desired path, the best precision one can achieve will be limited by the screening length whose value could be of the order of few nanometers in a 2DEG~\cite{datta1997electronic}.\\
Now regarding point (2), the parameter $\theta$, which parametrizes strength of the spin mixing [$\vert B_j\vert\equiv\sin(\theta/2)$], crucially depends on $\Gamma_j^2/(\Delta^2+{\Gamma_j}^2)$ which is small ($\sim0.13$) for realistic values of $\Gamma_j$ and $\Delta$~\cite{karmakar2011controlled} and it does not allow of full variation of $\theta$. This fact poses a hurdles for our  proposal. But this amplitude can be resonantly enhanced if a multiple path geometry as shown in lower inset of Fig.~\ref{fig:magprof}(b) is used. This essentially produces an effective multiple magnetic barrier problem where resonance can be tuned by tuning the distance between the barriers. Hence, it corresponds to a situation where the effective magnetic field profile on the edge is given by $\alpha(x)=\sum_{n=1}^N\{\Theta(x + l/2 - (n-1)\overline{\lambda+l})-\Theta(x - l/2 - (n-1)\overline{\lambda+l})\}$ with $\lambda$ being the effective spacing between consecutive magnetic patches and N is the number of such patches. The resonance condition is tuned simply by setting $\lambda=\lambda_{\rm res}=\pi v_{\rm F}/\Delta$ [see Fig.~\ref{fig:magprof}(c)] which reduces the effective scattering matrix for the N-patch to $S_j^N$.  This scattering matrix ($S_j^N$) has two very important properties both of which are crucial for our proposal to work. Firstly, the off-diagonal element of this matrix still has a neat $\phi_j$ dependence as it was for $S_j$ itself [see inset (ii) of Fig.~\ref{fig:magprof}(c)] and secondly, absolute square of the off-diagonal element ($t_{\uparrow\downarrow}$) of this matrix scales monotonically with $N$ [see inset (i) of Fig.~\ref{fig:magprof}(c)], hence, providing a control parameter for effective $\theta$ corresponding to the scattering matrix $S_j^N$. This concludes our feasibility study. \\
\noindent
{\underline{\sl{Interaction effects:-}}} Inter-edge coulomb interaction can play an important role in the mixer regions of our proposed set up where the copropagating edges come in close vicinity. Using the technique of bosonization to exactly diagonalize the inter-edge interaction leads to two new eigenmodes;  a slow neutral mode and a fast charge mode propagating with different renormalized velocities $v_{-}=v-g$ and $v_{+}=v+g$ respectively where $g$ is the strength of the local density-density type inter-edge interaction and $v$ is renormalized Fermi velocity due to intra-edge interaction~\cite{chirolli2013interactions}. It was shown in Ref. \onlinecite{chirolli2013interactions} that the contributions of these new modes in mixer region result in a bias and interaction dependent shift of the condition for resonance tunneling between the spin up and down edge modes given by $k_{\uparrow} - k_{\downarrow}=2\pi/\lambda +  2 g  eV/(v_+v_-)=2\pi/\lambda + e V/{(N \epsilon_c)}$ where $\epsilon_c=1/[N\lambda(1/{v_-}-1/{v_+})]$ is the relevant energy scale for an array of $N$ magnets forming the mixer. As along as  $eV \ll N \epsilon_c$, one can safely neglect the interaction effects ($\epsilon_c \approx 0.16 \mu eV$ for $N=10$ and $g=0.1v_{\rm F}$). Away from the mixers, the edges are spatially well separated from each other [see Fig.~\ref{fig:magprof}(a)] and hence interaction effects are heavily suppressed there. This clearly implies that for weak interaction ($g=0.1v_{\rm F}$) and within a bias window of the order of $N \epsilon_c \approx 1.6 \mu eV$ (a typical value used in noise measurements~\cite{neder2007interference}) for $N=10$, one can safely ignore effects due to such electronic interactions.\\
\noindent
{\underline{\sl{Conclusions :-}}} Our proposal pertains to electronic HB-T interferometer realized on a $\nu=2$ QH edge which is a  QH analog of ``intensity interferometry with polarized light". We show that pure spin manipulations can lead to oscillations solely in the cross-correlated noise and provide an efficient control of orbital entanglement production and detection. 
\\ 
\noindent
{\underline{\sl{Acknowledgments :-}}} We thank Moty Heiblum for valuable discussions and Mandar M. Deshmukh and Sthitadhi Roy for useful comments on the manuscript. PM would like to thank Joseph Samuel and Supurna Sinha for discussions on intensity interferometry. DW acknowledges financial support from the University Teaching Assistantship at the University of Delhi. PM acknowledges support from University Grants Commission under the second phase of University with Potential of Excellence at Jawaharlal Nehru University. SD acknowledges support from University of Delhi in the from of a research grant (RC/2014/6820).


\bibliography{reference}
\end{document}